\documentclass[aps,pre,preprint,groupedaddress,showpacs]{revtex4-1}
\usepackage{graphicx}
\usepackage{amsmath}
\begin{document}
\title{Exchange of helicity in a knotted electromagnetic field}
\author{Manuel Array\'as and Jos\'e L. Trueba}
\affiliation{\'Area de Electromagnetismo, Universidad Rey Juan
Carlos, Camino del Molino s/n, 28943 Fuenlabrada, Madrid, Spain}

\begin{abstract}
 In this work we present for the first time an
exact solution of Maxwell equations in vacuum, having non trivial
topology, in which there is an exchange of helicity between the
electric and magnetic part of such field. We calculate the temporal
evolution of the magnetic and electric helicities, and explain the
exchange of helicity making use of the Chern-Simon form. We also have
found and explained that, as time goes to infinity, both helicities reach the same value
and the exchange between the magnetic and electric part of the field stops.
\end{abstract}

\date{\today}
\maketitle

Topologically nontrivial solutions of Maxwell equations in vacuum have
been recently worth of some interest \cite{Arr10}. The possibility of
the experimental observation of some particular solutions with certain degree
of linkage in the field lines, called Ra\~nada electromagnetic knots 
\cite{Ran89,Ran92,Ran95,Ran97,Ran98} have been reported \cite{Irv08}. 
Such investigations are also relevant from the
fundamental point of view. Topologically nontrivial
solutions may open a new path to explore the stability of
electromagnetic fields, and the chance that a classical
field theory might serve as a model for stable elementary particles,
and idea investigated more than a century ago by Kelvin \cite{Kelvin} and
later by Wheeler using the concept of geon \cite{Wheeler}. In this work 
we present for the first time an exact solution of Maxwell equations in vacuum, 
having non trivial topology, in which there is an exchange of helicity between the
electric and magnetic parts of such field. The magnetic helicity
\cite{Ricca92,Berg99} can be defined as the average of the Gauss linking integral 
over all pairs of magnetic lines plus the self-linking number over all magnetic 
lines, so that it is a mean value of the linkage of the magnetic lines. Similarly,
for electromagnetic fields in vacuum the electric helicity can also be defined, that 
is a mean value of the linkage of the electric lines. We have calculated 
the temporal evolution of the magnetic and electric helicities, and explained the
exchange of helicity making use of the Chern-Simon form. We have
found and proved that as time goes to infinity, both helicities reach the same value
and the exchange between the magnetic and electric part of the field stops. We also 
have given some explicit cases in which there is linkage of field lines but the helicity is zero, 
suggesting that the relation between helicity and field line topology needs to be further
investigated. 

We begin with the computation, at time $t=0$, of a magnetic field in which
any pair of magnetic lines is linked, with a linking number equal to 1, and an
electric field, at $t=0$, in which there is no linkage. In order to find these
fields we use the method described in \cite{Ran01} to find electromagnetic knots 
(although the electromagnetic field that we will use in the present work 
{\it is not} a Ra\~nada electromagnetic knot). Let $\phi ({\bf r})$, $\theta ({\bf
r})$, two complex scalar fields such that can be considered as maps
$\phi, \theta : S^3 \rightarrow S^2$ after identifying the
physical space $R^3$ with $S^3$ and the complex plane with $S^2$. Doing so, it is assumed 
that both scalars have only one value at infinity. We will impose that at 
$t=0$ the level curves of the scalar fields $(\phi, \theta)$, coincide with 
the magnetic and electric lines respectively, each one of these lines being labelled 
by the constant value of the corresponding scalar. Given $\phi$ and $\theta$ with 
these conditions, we can construct the magnetic and electric fields at $t=0$ as 
\begin{eqnarray} 
{\bf B}({\bf r},0) = \frac{\sqrt{a}}{2\pi
i}\frac{ \nabla \phi \times \nabla {\bar{\phi}}}{(1+
{\bar{\phi}}\phi )^{2}},  \nonumber \\
{\bf E}({\bf r},0) = \frac{\sqrt{a} c}{2\pi
i}\frac{\nabla {\bar{\theta}}\times
\nabla \theta}{(1+{\bar{\theta}}\theta)^{2}}, \label{11.3}
\end{eqnarray}
where ${\bar{\phi}}$ and ${\bar{\theta}}$ are the complex
conjugates of $\phi$ and $\theta$ respectively, $i$ is the
imaginary unit, $a$ is a constant introduced so that the magnetic and electric fields have correct
dimensions, and $c$ is the velocity of light in vacuum. In the SI of units, 
$a$ can be expressed as a pure number times the Planck constant $\hbar$ times the light 
velocity $c$ times the vacuum permeability $\mu_{0}$. 

\begin{figure}
\centering
\includegraphics[width=0.35\textwidth]{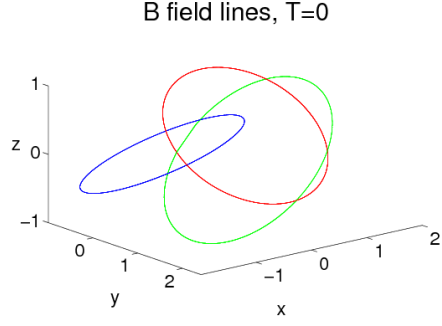}
\caption{Some magnetic field lines at time $t=0$. These lines coincide with level curves of the Hopf 
map $\phi$ (\ref{12.1}) due to the construction given by the first equation in (\ref{11.3}). This means that all 
the lines are linked to each other with a linking number equal to 1. The electric field is zero in all 
the points at $t=0$.} 
\label{fig1}
\end{figure}

The construction given in equations (\ref{11.3})
assures that all pairs of lines of the field ${\bf B} ({\bf r},0)$ are linked, and that the linking
number is the same for all the pairs and it is given by the Hopf index \cite{Hopf} of the
map $\phi$. Similarly, all pairs of lines of the field ${\bf E} ({\bf r},0)$ are linked, the linking
number of all pairs of lines is the same and it is given by the Hopf index of the
map $\theta$. This means that, at $t=0$, the linkage of the magnetic and the electric lines
is set, the initial magnetic helicity is proportional to the Hopf index of
$\phi$ and the initial electric helicity is proportional to the Hopf index of
$\theta$. It is convenient to work with dimensionless coordinates in the mathematical spacetime 
$S^3 \times R$, and in $S^2$. In order to do that, we define the dimensionless coordinates 
$(X, Y, Z, T)$, related to the physical ones $(x, y, z, t)$ (in the SI of units that we will use 
in this work) by $(X, Y, Z, T) = (x, y, z, c t)/L_{0}$, and $r^2 /L_{0}^2 =(x^2 + y^2 +z^2)/L_{0}^2 
= X^2 + Y^2 +Z^2 =R^2$, where $L_{0}$ is a constant with dimensions of length. Now, we will 
choose for the scalar fields the Hopf map and the zero map,
\begin{equation}
\phi =\frac{2(X+iY)}{2Z+i(R^{2}-1)} , \, \, \theta =0.
\label{12.1}
\end{equation}
In Fig.~\ref{fig1} we can see that the lines at $t=0$ of the magnetic field constructed in this way are 
linked to each other, the linking number being 1 in all the cases. Obviously, since $\theta=0$, the 
initial electric field is zero at every point.

To find the electromagnetic field at any time from the Cauchy data (\ref{11.3}), with $\phi$ and $\theta$
given by expressions (\ref{12.1}), we use Fourier analysis. The fields turn out to be,
\begin{eqnarray}
{\bf B}({\bf r},t) = \frac{\sqrt{a}}{\pi L_{0}^2(A^{2}+T^{2})^{3}}
\left( Q{\bf H}_{1}+P{\bf H}_{2}\right) ,  \nonumber \\
{\bf E}({\bf r},t) = \frac{\sqrt{a} c}{\pi L_{0}^2
(A^{2}+T^{2})^{3}}\left( Q{\bf H}_{2}-P{\bf H}_{1}\right) ,
\label{11.35}
\end{eqnarray}
where the quantities $A$, $P$, $Q$ are defined by $A=( R^{2}-T^{2}+1)/2$, $P=T(T^{2}-3A^{2})$, 
$Q=A(A^{2}-3T^{2})$, and the vectors ${\bf H}_{1}$ and ${\bf H}_{2}$ are
\begin{eqnarray}
{\bf H}_{1}&=& \left( Y-XZ,-X-YZ,\frac{-1-Z^{2}+X^{2}+Y^{2}+T^{2}}{2}%
\right)  , \nonumber \\
{\bf H}_{2}&=&\left( -TY, TX, T \right)  . \label{11.37}
\end{eqnarray}
The magnetic and electric helicities of an electromagnetic field in vacuum, in the SI of units, can be defined as
\begin{equation}
h_{m}=\frac{1}{2 \mu_{0}} \int_{R^{3}}{\bf A}\cdot {\bf B}\,d^{3}r, \, \, h_{e}=
\frac{\varepsilon_{0}}{2} \int_{R^{3}}{\bf C}\cdot {\bf E}\,d^{3}r, \label{11.29}
\end{equation}
where $\varepsilon_{0}$ is the vacuum permittivity, and {\bf A} and {\bf C} are the potential vectors 
given by the conditions that
\begin{equation}
{\bf B}=\nabla\times{\bf A}, \, \, {\bf E}=\nabla\times{\bf C} \label{vectorp}.
\end{equation}
Since the Hopf index of the map $\phi$ is 1, and the Hopf index of the map $\theta$ is 0, at $t=0$ 
the magnetic and electric helicities of the knotted electromagnetic field given by equations (\ref{11.35}) are
\begin{equation}
h_{m} (0) =  \frac{a}{2 \mu_{0}}  , \, \, h_{e} (0)  = 0. \label{11.299}
\end{equation}
An important quantity related to helicities in any electromagnetic field in vacuum is the electromagnetic helicity 
$h = h_{m} + h_{e}$, that is the sum of the magnetic and electric helicities of the field and is a constant in 
the evolution of the electromagnetic field. In the case of the knotted electromagnetic field used in this work, 
$h = a/(2 \mu_{0})$. Since $h$ is constant in the evolution of the field, we may expect that the electromagnetic field we have obtained will retain certain linkage and remain topologically nontrivial during its time evolution. 

\begin{figure}
\centering
\includegraphics[width=0.45\textwidth]{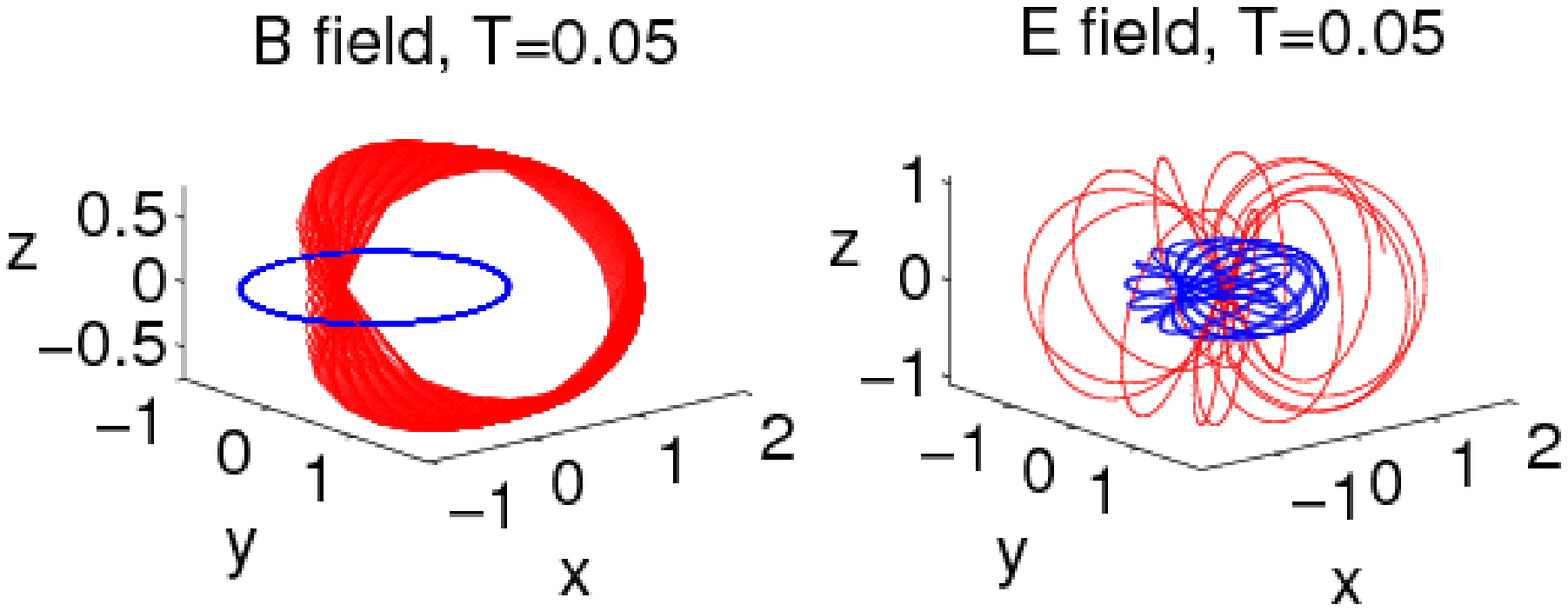}
\includegraphics[width=0.45\textwidth]{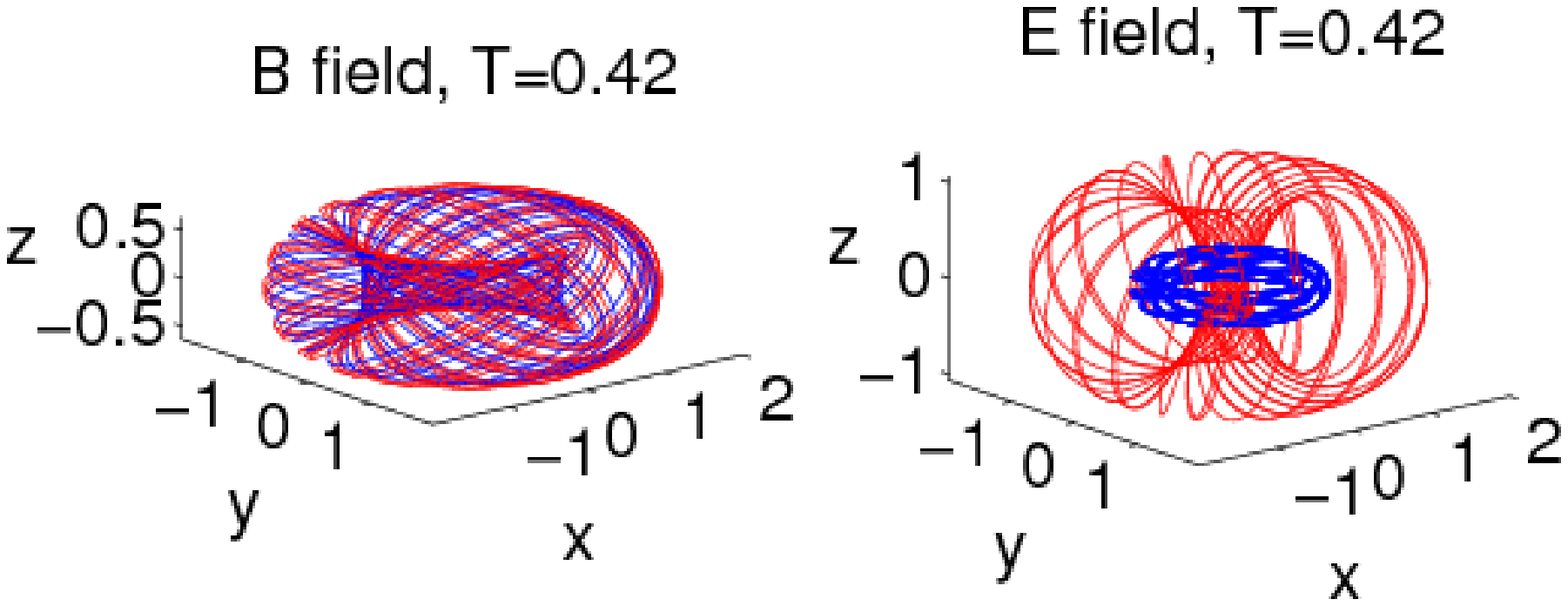}
\includegraphics[width=0.45\textwidth]{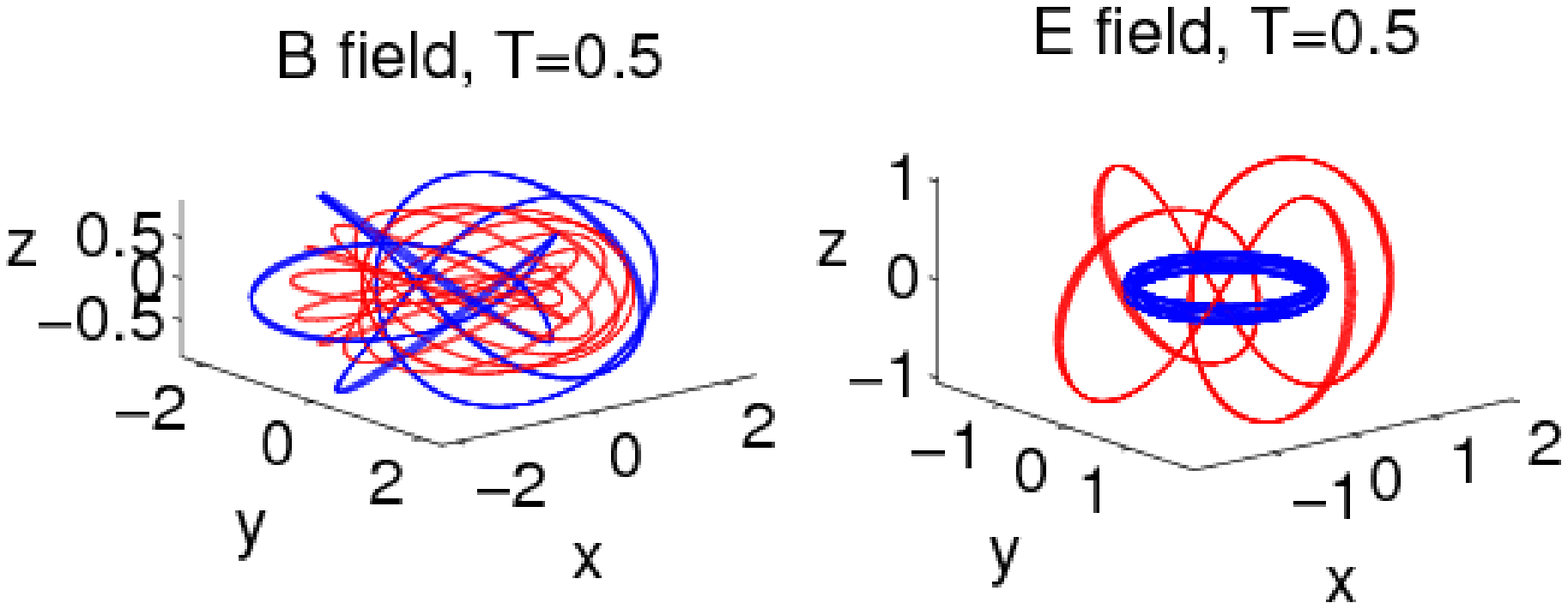}
\includegraphics[width=0.45\textwidth]{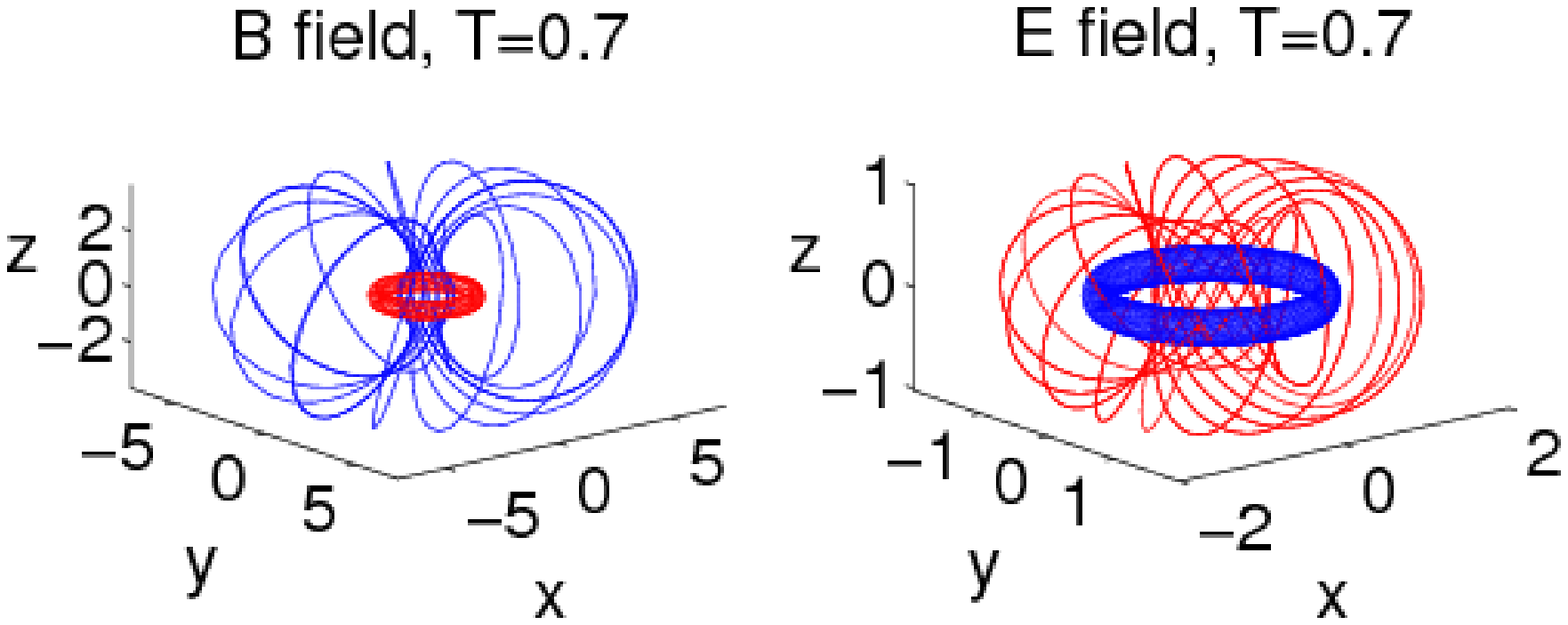}
\caption{Time evolution of field lines for different times. By looking at Fig.~\ref{fig3}, we note that 
at instant $T=0.42$ it is satisfied that $h_m=h_e$. Moreover, at $T=0.7$, $h_m$ reaches its minimum and 
$h_e$ its maximum. For all the times the numerical integration results in some closed field lines while
others are not closed.} 
\label{fig2}
\end{figure}

\begin{figure}
\centering
\includegraphics[width=0.45\textwidth]{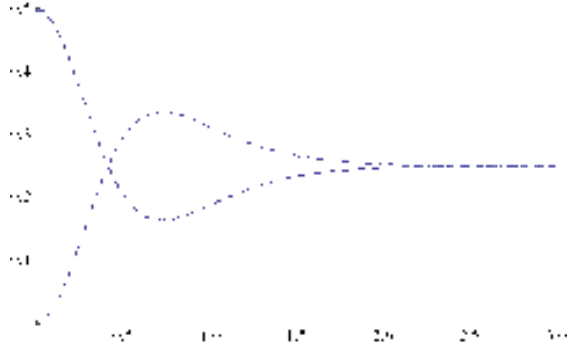}
\caption{Time evolution of the magnetic $h_m$ (dashed line) and electric $h_e$ (dot-dashed line) helicities 
in units of $a/\mu_0$. The sum of both is the total constant electromagnetic helicity (straight line), that
is equal to $1/2$ in the same units. The interchange between magnetic and electric helicities is clear in the
figure. Note also how both helicities reach a common value $h_{m}=h_{e}=h/2$ for large values
of time.} 
\label{fig3}
\end{figure}

In Fig.~\ref{fig2} we have numerically integrated the equations for some magnetic and electric lines 
to see how they evolve in time. Note as the linkage is clear for both magnetic and electric lines 
(although the electric field was initially zero). The behavior of the field lines remains topologically 
nontrivial, meanwhile there must be an interchange of helicity between the magnetic and electric parts 
of the electromagnetic field. Initially all the magnetic lines are linked to each other. As time evolves, 
not all the lines will remain linked, but certainly there are linked ones. On the other hand, we can 
observed that being initially the electric field equal to zero, as we move on in time the topology 
of the electric field lines seems to follow the magnetic one. In the pictures, for the instant
$T=0.42$, the magnetic helicity $h_m$ and the electric helicity $h_e$ become equals, and at 
$T=0.7$ the magnetic helicity $h_m$ reaches its minimum and $h_e$ its maximum (see also 
Fig.~\ref{fig3} to understand what goes on with the helicities for the instants of time chosen).
An important feature obtained in the numerical computations is that, for any time $T>0$, one can find
closed and unclosed field lines (both electric and magnetic). Since at $T=0$ all the lines are closed,
the mechanism that give rise to open lines for $T>0$ is not completely clear and even could be a numerical
artifact. This question deserves a further study. 

To investigate the interchange of helicity between the electric and magnetic fields we use the concept 
of helicity density 4-current or Chern-Simons form (see \cite{Hehl03} for a nice explanation). The 
magnetic helicity density is the zeroth component of the Chern-Simons 4-current
\begin{equation}
{\cal H}_{m}^{\mu} =  \frac{1}{2 \mu_{0}} \, A_{\nu} G^{\nu \mu}.
\label{12.2}
\end{equation}
Here, $\mu, \nu = 0, 1, 2, 3$ and $i, j, = 1, 2, 3$, and $A^{\mu} = (V/c, {\bf A})$ is the 4-vector potential 
of the electromagnetic field (remember that we use SI units in this work), so that the electromagnetic tensor is 
$F_{\mu \nu} = \partial_{\mu} A_{\nu} - \partial_{\nu} A_{\mu}$. From this tensor one finds the components
of the electric field as ${\bf E}_{i} = c\, F^{i0}$, and the magnetic field as 
${\bf B}_{i} = -\epsilon_{ijk} F^{jk}/2$ as usual. Moreover, 
\begin{equation}
G_{\mu \nu} = \frac{1}{2} \, \epsilon_{\mu \nu \alpha \beta} F^{\alpha \beta}, 
\label{dual}
\end{equation}
is dual to the electromagnetic field $F^{\mu \nu}$, with components ${\bf B}_{i} = G^{0i}$, 
${\bf E}_{i} = - c\, \epsilon_{ijk} G^{jk}/2$. As long as we are studying fields in vacuum, we can define
another 4-vector potential $C^{\mu} = (C^{0}, {\bf C})$, so that $G_{\mu \nu} = \left( \partial_{\mu} C_{\nu} 
- \partial_{\nu} C_{\mu} \right)/c$, and we can take the electric helicity density to be the zeroth component 
of the 4-current (see \cite{Ran01}),
\begin{equation}
{\cal H}_{e}^{\mu} = - \frac{1}{2 c \mu_{0}} \, C_{\nu} F^{\nu \mu}.
\label{12.3}
\end{equation}
The divergence of ${\cal H}_{m}^{\mu}$ and ${\cal H}_{e}^{\mu}$ gives the conservation law for both 
helicities. It turns out that
\begin{equation}
\partial_{\mu} {\cal H}_{m}^{\mu} = - \frac{1}{4 \mu_{0}} \, F_{\mu \nu} G^{\mu \nu} , \, \, \partial_{\mu} {\cal H}_{e}^{\mu} = 
\frac{1}{4 \mu_{0}} G_{\mu \nu} F^{\mu \nu}.
\label{12.3a}
\end{equation}
From this equation, by integrating in $R^3$, the time derivatives of the magnetic and electric helicities are obtained as
\begin{eqnarray}
\frac{dh_{m}}{dt} &=& c \int_{R^{3}} \partial_{0} {\cal H}_{m}^{0} \,d^{3}r =  - \frac{1}{\mu_{0}} \int_{R^{3}} 
{\bf E} \cdot {\bf B} \,d^{3}r, \nonumber \\  
\frac{dh_{e}}{dt} &=& c \int_{R^{3}} \partial_{0} {\cal H}_{e}^{0} \,d^{3}r =  \frac{1}{\mu_{0}} \int_{R^{3}} 
{\bf E} \cdot {\bf B} \,d^{3}r.
\label{12.4}
\end{eqnarray}

Note that equations (\ref{12.2}) -- (\ref{12.4}) are general results for electromagnetic fields in vacuum.
From these equations one can extract some consequences. First, let us consider the case in which the spatial
integral of the Lorentz invariant quantity ${\bf E} \cdot {\bf B}$ for a given electromagnetic field 
in vacuum is zero. In this case both the magnetic and the electric helicities are constant during the
evolution of the field. Since these quantities are related to the Gauss linking integral and the self-linking
number of the magnetic and the electric lines, respectively, if they are not zero then one can expect that there 
will be a certain degree of linkage in the lines and this linkage will not disappear in time. Nice particular
examples of this case are Ra\~nada electromagnetic knots \cite{Arr10,Ran89,Ran92,Ran95,Ran97,Ran98,Irv08,Ran01} 
in which, by construction, always satisfy that the electric and magnetic fields are mutually orthogonal. Moreover, 
in the case of Ra\~nada electromagnetic knots the magnetic helicity is equal to a topological invariant, the Hopf
index of a map between the compactified space $R^3$ and the compactified complex plane, and the electric helicity 
is equal to the Hopf index of another map between the compactified space $R^3$ and the compactified complex
plane. Since these helicities are conserved during time evolution of
the electromagnetic field, it is concluded that the topology of the
field lines (magnetic and electric) is conserved and characterized by
the value of the helicities. Because in this case, the Hopf index $n$
is related to the Gauss linking number $\ell$ by $n=\ell \mu ^2$,
being $\mu$ is the multiplicity of the level curves (i.e. the number
of different magnetic or electric lines that have the same label
$\phi$ or $\theta$), all the field lines remains linked during the
time evolution (a picture of this case would be the situation shown in Fig.~\ref{fig1}, 
with the linkage maintained in time).

However, if ${\bf E} \cdot {\bf B} \neq 0$ as in the case of the knotted 
electromagnetic field presented in this work, then the time derivatives of the 
magnetic and electric helicities are not zero but from equation (\ref{12.4}) 
one can see that they satisfy $d h_{m}/dt = - d h_{e}/dt$,
so that there is an interchange of helicities between the magnetic and
electric parts of the field. In the case of the knotted
electromagnetic field presented here, it is only at $t=0$ that the
magnetic helicity is related to the Hopf index of the map $\phi$
between the compactified $R^3$ and the compactified complex plane. We also recall that the
electric helicity at $t=0$ is zero. This is why the relation between topology of lines
and helicities is not so clear in this case. We have computed explicitly, using equation 
(\ref{12.4}), how both helicities behave for any value of time while the electromagnetic
field varies. The result can be seen in Fig.~\ref{fig3}. One can see in this figure
that helicities are transferred between the magnetic and the
electric parts of the field. How this interchange is related to a
change in the topology of the field lines is not clear since, for
$t>0$, it is not possible to relate magnetic and/or electric helicity
to a topological invariant as the Hopf index of any map. However, it
seems from numerical computations (see Figs.~\ref{fig1} and \ref{fig2}), that the
topology of both the magnetic and electric field lines is similar for
$t>0$, and that a nonzero initial helicity results in a nontrivial
topology of the field lines for any time.

Independently of the value of the Lorentz invariant quantity ${\bf E} \cdot {\bf B}$, the electromagnetic 
helicity $h$, whose density 4-current is given by
\begin{equation}
{\cal H}^{\mu} = \frac{1}{2 \mu_{0}} \, A_{\nu} G^{\nu \mu} - \frac{1}{2 c \mu_{0}} C_{\nu} F^{\nu \mu},
\label{12.5}
\end{equation}
is a conserved quantity for any electromagnetic field in vacuum (and it is the quantity related to 
the photon content of the field as we discuss below).
In Fig.~\ref{fig3} we have plotted the value of the electromagnetic helicity $h=h_{m}+h_{e}$ 
and we can see that it is constant indeed. This conservation also affects to the value of magnetic and 
electric helicities for large values of time. Magnetic and electric helicities can be written, 
respectively, as $h_{m}=(h + \tilde{h} )/2$, $h_{e}=(h - \tilde{h} )/2$, where $h$ is the
electromagnetic helicity and $\tilde{h}$ is the difference between the magnetic and the electric 
helicities. If the vector potentials ${\bf A}$ and ${\bf C}$ and the fields ${\bf B}$ and ${\bf E}$
are such that 
\begin{equation}
\lim_{t \rightarrow \infty} \int_{R^3} \left( {\bf A} \cdot {\bf B} - \frac{1}{c^2} \, {\bf C} 
\cdot {\bf E} \right) \, d^3 r = 0, \label{difference}
\end{equation}
then, when $t$ is large, both helicities coincide, i. e. $\lim_{t \rightarrow \infty} h_{m} (t) = 
\lim_{t \rightarrow \infty} h_{e} (t)$. This result suggests that only in the case that initial 
helicities are different and (\ref{difference}) is fulfilled, they are going to exchange to one 
another in order to relax to a common value. This is the case that we have presented here.

The exchange of helicities between the magnetic and electric parts of a knotted 
electromagnetic field in vacuum reported in this work leave one question to be answered: 
How can we characterize the topology of the field lines? The answer may have some 
important physical consequences since there is a relation between the electromagnetic helicity
and the classical expression for the difference of right- and left-handed photons of the
electromagnetic field \cite{Ran97}. According to this relation, and taking into account the 
results of Ref. \cite{Ricca92}, adding a right  or left photon would imply to add or remove 
crossings of the field lines. In order to investigate such questions, the explicit solutions 
of Maxwell equations given by the expressions (\ref{11.35}) may play a central role. These 
solutions allow helicity exchange between the magnetic and electric parts of an electromagnetic 
field in vacuum but, at the same time, the topology of lines remains being not trivial during
the evolution of the field.  

The authors thank very useful discussions on parts of this work with Antonio F. Ra\~nada, 
Jos\'e M. Montesinos and Daniel Peralta-Salas. This work has been supported by the Spanish 
Ministerio de Ciencia e Innovaci\'on under project AYA2009-14027-C07-04.

\end{document}